\documentclass{aa}
\usepackage{graphicx,latexsym}

\def\cm-2{cm$^{-2}$}

\def\chandra{{\it Chandra}}
\def\ngc253{\object{NGC~253}}
\def\m82{\object{M82}}
\def\casa{\object{Cas\,A}}
\def\n132d{\object{N\,132D}}

\begin{document}
 
   \title{XMM-Newton observations of \ngc253: Resolving the emission 
          components in the disk and nuclear area\thanks{Based 
    on observations with XMM-Newton, an ESA Science Mission 
    with instruments and contributions directly funded by ESA Member
    States and the USA (NASA).}}

   \author{W.~Pietsch\inst{1} \and
           T.P.~Roberts\inst{2} \and
           M.~Sako\inst{3} \and
           M.J.~Freyberg\inst{1} \and
           A.M.~Read\inst{1} \and
           K.N.~Borozdin\inst{4} \and
           G.~Branduardi-Raymont\inst{5} \and
           M.~Cappi\inst{6} \and
           M.~Ehle\inst{7}  \and
           P.~Ferrando\inst{8} \and
           S.M.~Kahn\inst{3} \and
           T.J.~Ponman\inst{9} \and
           A.~Ptak\inst{10} \and
           R.E.~Shirey\inst{11} \and
           M.~Ward\inst{2}
        }
   \offprints{W.~Pietsch, \email{wnp@mpe.mpg.de}}

   \institute{Max Planck Institut f\"ur extraterrestrische Physik,
              Giessenbachstra\ss e, D-85741 Garching, Germany
   \and       Department of Physics \& Astronomy, University of Leicester, Leicester LE1 7RH, UK
   \and       Columbia Astrophys. Lab. and Dep. of Phys., Columbia Univ.,
              550 W. 120th St., New York, NY 10027, USA
   \and       NIS-2, Space and Remote Sensing Sciences, MS D436 Los Alamos National Lab.,
              Los Alamos, NM 87545, USA
   \and       Mullard Space Science Lab., University College London,
              Holmbury St. Mary, Dorking, Surrey, RH5 6NT, UK
   \and       Istituto TeSRE/CNR, Via Gobetti 101, 40129 Bologna, Italy
   \and       XMM-Newton SOC, Villafranca,  Apartado 50727, E-28080 Madrid, Spain
   \and       DAPNIA/Service d'Astrophys., Bat. 709 Orme des Merisiers, CEA Saclay,
              91191 Gif-sur-Yvette, Cedex, France
   \and       School of Physics \& Astronomy, University Birmingham, Birmingham B15 2TT, UK
   \and       Department of Physics, Carnegie Mellon University, USA
   \and       Department of Physics, University of California, Santa Barbara, CA 93106, USA
        }          
   \date{Received date; accepted date}
   \titlerunning{XMM-Newton observations of \ngc253} 

\abstract{
We describe the first XMM-Newton observations of the starburst galaxy \ngc253.
As known from previous X-ray observations, \ngc253 shows a mixture of extended
(disk and halo) and point-source emission. The high XMM-Newton throughput allows 
a detailed investigation of the spatial, spectral and
variability properties of these components simultaneously. 
We characterize the brightest sources by their hardness ratios, detect a bright
X-ray transient $\sim$70\arcsec\ SSW of the nucleus, and show the spectrum and
light curve of the brightest point source ($\sim$30\arcsec\ S of the nucleus,
most likely a black-hole X-ray binary, BHXRB). The unresolved emission of two 
disk regions can be modeled by two thin thermal plasma components (temperatures
of $\sim$ 0.13 and 0.4 keV) plus residual harder emission, with the lower
temperature component originating from above the disk. The nuclear spectrum
can be modeled by a three temperature plasma ($\sim$ 0.6, 0.9,
and 6 keV) with the higher temperatures increasingly absorbed. The high temperature
component most likely originates from the starburst nucleus, as no non-thermal
component, that would point at a significant contribution from an
active nucleus (AGN), is needed.
Assuming that type\,{\rm II}a supernova remnants (SNRs) are
mostly responsible for the E$>$4 keV emission, the detection with EPIC of the 
6.7\,keV line allows us to estimate a  
supernova rate within the nuclear starburst of 0.2\,yr$^{-1}$.
The unprecedented combination
of RGS and EPIC %spectra and images in lines 
also sheds new light on the emission
of the complex nuclear region, the X-ray plume and the disk diffuse emission.
In particular, EPIC images reveal that the limb-brightening of the plume is 
mostly seen in higher ionization emission lines, while in the lower ionization 
lines, and below 0.5\,keV, the plume is more homogeneously structured.
The plume spectrum can again be modeled by a three temperature thermal plasma
containing the two low temperature nuclear components (though less absorbed) 
plus an unabsorbed 0.15~keV component similar to the disk spectra. 
This points to new interpretations as to the make up of the starburst-driven 
outflow. 
\keywords{X-rays: galaxies -- Galaxies: individual: \ngc253 --
                 Galaxies: spiral  -- Galaxies: starburst -- 
                 Interstellar medium: jets and outflows}}
\maketitle

\section{Introduction}

X-ray emission from starburst galaxies is known to be complex, revealing 
both point sources and diffuse X-ray emission in abundance. 
X-ray binaries, supernovae, supernova remnants and nuclear sources
dominate the point-like source contribution, while the hot phases of the
interstellar medium (ISM), in the form of hot outflows (or winds), more bound
coronal features and diffuse  emission within the disk, make up the
diffuse, gaseous component. 

The nearby edge-on galaxy \ngc253 is perhaps the classic example (along with
\m82) of a starburst galaxy, and as such, has received a great deal of attention
from X-ray observatories over the years. Initial {\em Einstein} observations
(Fabbiano \& Trinchieri \cite{fab84}; Fabbiano \cite{fab88}), along with seeing several bright
point-like disk sources, discovered large plumes of diffuse emission extending
above and below the disk of the galaxy.  This emission is thought to be due to 
``mass-loading" of a hotter wind with cooler
ambient gas, i.e. shock heated and accelerated cooler interstellar and circumnuclear 
gas (Heckman et al. \cite{hec90}; Strickland et al. \cite{str00}).

Several studies were made of \ngc253 with ROSAT. Both the PSPC and HRI
data were presented by Read et al. (\cite{rea97}) and Dahlem et al. (\cite{dahlem}). 
An extensive ROSAT point source
catalogue of \ngc253 by Vogler \& Pietsch (\cite{vog99}) made it possible to separate
the point source and diffuse X-ray emission, allowing insights into the spatial,
spectral and timing properties of the many point sources within \ngc253. The
`nuclear', likely starburst-associated source appeared to be extended, and the
brightest point-like source, lying some ~30\arcsec\ south of the nucleus, at the
border of a plume of diffuse emission, was thought to be a plausible 
BHXRB candidate. Structure in the diffuse emission could now be studied in
detail, Pietsch et al. (\cite{pie00}) reporting different diffuse emission components in
the nucleus, disk and halo. Especially of note was the discovery of coronal
diffuse emission bubbling out of the disk, via galactic {\em fountains} and {\em
chimneys} (essentially formed by localized high-activity star-forming regions
within the disk), then falling ballistically back to the plane (e.g. Norman \&
Ikeuchi \cite{nor89}). Also observed was a hollow-cone shaped diffuse plume of emission 
extending up to $\sim$700\,pc along the SE minor axis, thought to be due to the 
interaction of the galactic superwind and the dense disk ISM. 

More recent high spectral resolution observations with ASCA (Ptak et al. \cite{pta97})
have revealed strong O, Ne, Fe, Mg, S and Si emission lines in the integrated
\ngc253 spectrum, and these results have been backed up by BeppoSAX observations
(Cappi et al. \cite{cappi}), where for the first time, the 6.7\,keV Fe K line has been
detected (Persic et al. \cite{per98}). 
\chandra\ observations allowed for the first time a detailed study of 
the SE plume, tentatively examined with ROSAT, and showed it 
to be a well-collimated, limb-brightened kpc-scale conical outflow, which closely 
follows in morphology the known H$\alpha$ outflow. 
Furthermore, several point 
sources are seen, the nuclear feature being partly separated into a number of distinct 
sources (Strickland et al. \cite{str00}). 

In this letter we report the results of the performance verification
phase observation of \ngc253 to demonstrate XMM-Newton's capabilities of spatially
resolved spectroscopy in a mixture of point sources and diffuse emission.

\section{Observations and data analysis}

\begin{table}
      \caption{Details of \ngc253 XMM-Newton exposures. Exposure number (E), 
       instrument setting and exposure duration and low background time (LB) 
       are given.}
         \label{expo}
         \begin{flushleft}
         \begin{tabular}{lllllr}
            \hline
            \noalign{\smallskip}
E   & Det. &Filter &Mode &Dur. &LB\\
   &      &       &     &(s)&(s)\\
            \noalign{\smallskip}
            \hline
            \noalign{\smallskip}
1
 & RGS1  & &SPECTR + Q & 60\,613 & 47\,313\\
 & RGS2  & &SPECTR + Q & 60\,593 & 47\,293\\
 & MOS1  &Medium& PRI FULL & 38\,498 & 34\,492\\
 & MOS2  &Thin  & PRI FULL & 38\,498 & 34\,489\\
 & PN    &Medium& PRI FULL & 39\,000 & 35\,767\\
\noalign{\smallskip}
2
 & RGS1  & &SPECTR + Q & 17\,402 & 6\,102\\
 & RGS2  & &SPECTR + Q & 17\,406 & 6\,106\\
 & MOS1  &Medium& PRI FULL & 13\,593 &  7\,190\\
 & MOS2  &Thin  & PRI FULL & 13\,597 &  7\,188\\
 & PN    &Thin  & PRI FULL & 13\,999 &  7\,400\\
\noalign{\smallskip}
\hline
         \end{tabular}
         \end{flushleft}
   \end{table}

\begin{figure*}
  \resizebox{12cm}{!}{\includegraphics[bb=43 46 559 500,clip]{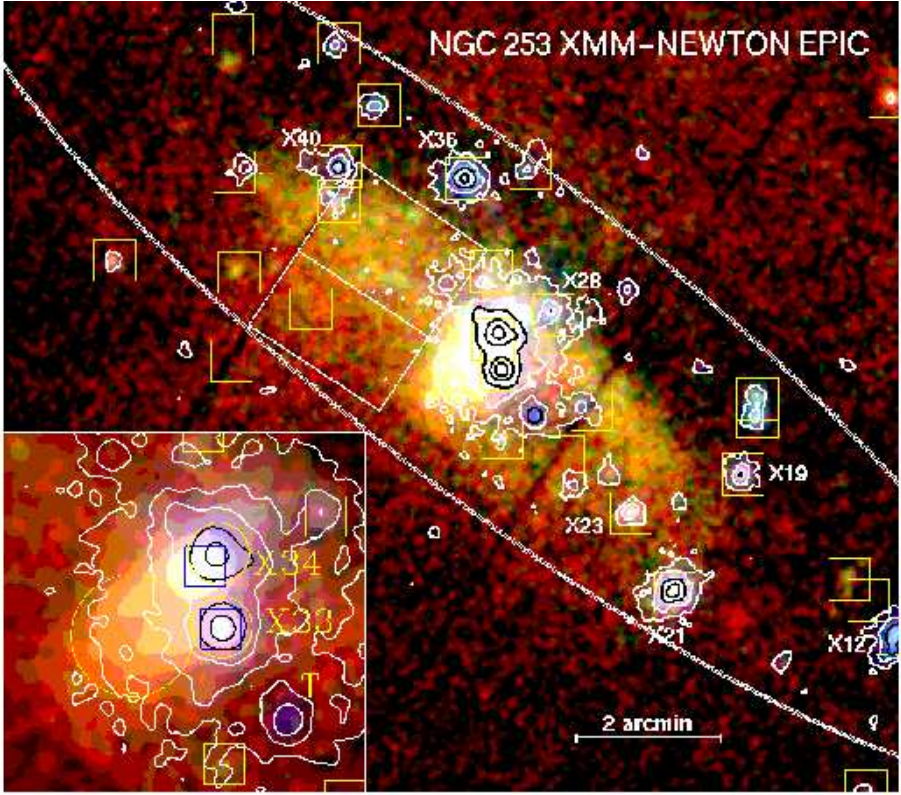}}
  \hfill
  \parbox[b]{55mm}{
    \caption[]{
Logarithmically-scaled, three-colour XMM-Newton EPIC image of the \ngc253 disk
and nuclear regions. Emission detected in all three EPIC instruments (MOS1,
MOS2 and PN) has been separately cleaned and merged together. Red, green and
blue show respectively the ROSAT-equivalent (0.2--0.5)~keV, (0.5--0.9)~keV 
and (0.9--2.0)~keV bands, while the hard
(2--10)~keV emission is shown superimposed as black/white 
contours at levels increasing by factors of 3 from 0.3\,ct~arcsec$^{-2}$. 
The data in each energy band have been smoothed with a PSF-equivalent Gaussian
of FWHM 5\arcsec. Shown in detail to the lower left is a higher-threshold 
factor two zoom-in on the nuclear region. Squares indicate the
position of ROSAT-detected sources (Vogler \& Pietsch \cite{vog99}), and source
identifications referred to in this paper (including the newly discovered
transient, T) are marked. Spectral extraction regions (white rectangles and, in 
inset, yellow circles) and the inclination-corrected optical D$_{25}$ ellipse
of \ngc253 are marked
}
    \label{color}}
\end{figure*}

\ngc253 was observed with the European X-ray observatory XMM-Newton 
(Jansen et al. \cite{jan01}) during orbit 89 on July 3rd and 4th 2000 for two
exposures each of the
European Photon Imaging Camera instruments 
(EPIC, Turner et al. \cite{tur01}; Str\"uder et al. \cite{strue01}) 
and the Reflection Grating Spectrometer (RGS, den Herder
et al. \cite{her01}). For details of the instrument setups and exposure durations 
for the different instruments see Table~\ref{expo}.
The position angle of the observation (55\degr) 
was close to the position angle of the galaxy (52\degr), and so, to
minimize the effect of the EPIC PN CCD boundaries on the plume and disk
emission, the nominal on-axis pointing position
was offset by 65\arcsec\ to the SE of the \ngc253 nucleus along the 
galaxy's minor axis. 
  
The standard reduction of the EPIC and RGS data was performed using the
latest version of the
Science Analysis System (SAS). This involved the subtraction of hot,
dead or flickering pixels, removal of events due to electronic noise and, for
the EPIC detectors, correction of event energies for charge transfer losses.
Also, times of high background were excluded to maximize sensitivity to low
surface brightness emission. While exposure 1 was interrupted by just a short
background flare, a major part of exposure 2 suffered from high background
(see duration of low background (LB) per exposure in Table \ref{expo}). Source
searching was performed on the cleaned MOS1, MOS2 and PN datasets separately.
After comparing these source lists, the MOS datasets were transformed to the
PN positions, whereupon, merging of the cleaned, aligned EPIC datasets took
place. A final transformation of the EPIC merged dataset to ROSAT positions
was then performed.

For fitting PN spectra and for defining spectral model lines in the hardness
ratio colour-colour plots we used response matrices  
provided by the EPIC instrument team (PN\_jul00 and for MOS v3.14).

For the RGS, source events from the total observation were
first extracted using a 1\arcmin\ spatial mask in the cross-dispersion
direction (i.e. along the minor axis centered on the bright nuclear area), 
and subsequently with a dispersion/pulse-height mask to select
the first-order photons. The background spectrum was estimated using the
same observation from regions $>$ 1\arcmin\ from the nucleus, which may
contain events from the galaxy's diffuse emission. Data from 
the two RGS's are combined
and divided by the exposure and the effective area of the instrument.

\section{X-ray emission from the disk}

\begin{figure*}
  \resizebox{\hsize}{!}{\includegraphics[angle=-90,clip]{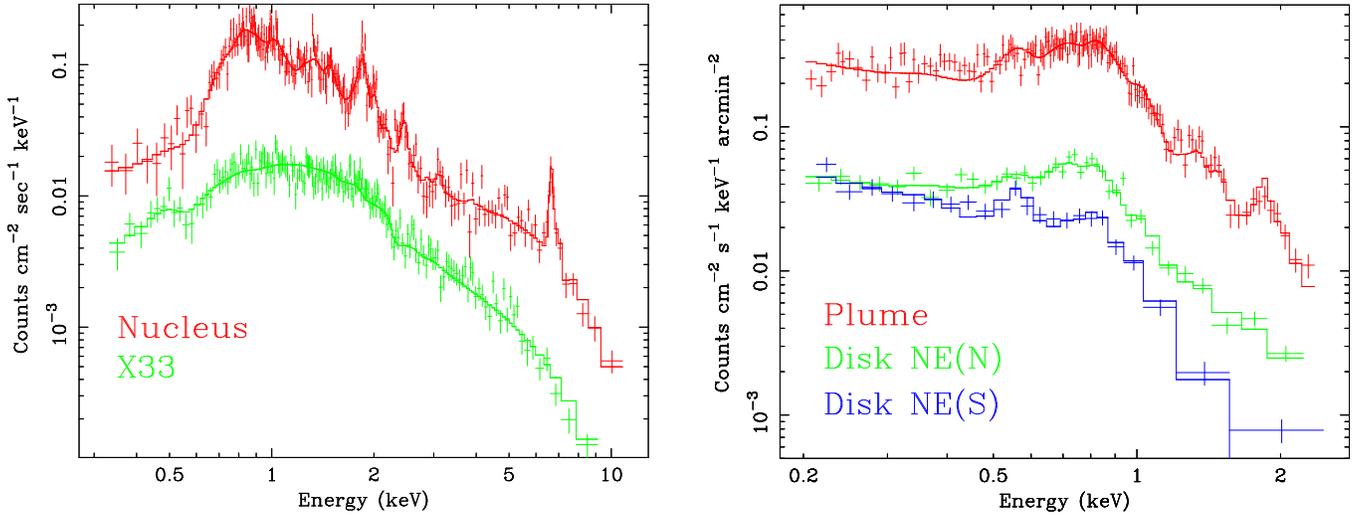}
                        \hspace{0.5cm}
                        \includegraphics[angle=-90,clip]{XMM18f2b.eps}}
    \caption[]{
EPIC PN background subtracted spectra from low background times of the first exposure with
spectral models (see text) indicated. {\bf Left:} Spectra of 
extended nucleus and X33 (extraction radius 12\farcs5, local background 
subtracted, X33 intensity shifted down by 0.5 decade for clarity). 
{\bf Right:} spectra of SE X-ray plume and two areas of the 
disk NE of the nucleus (0.5, 2.5, and 2.4 arcmin$^2$ extraction regions,
respectively [see Fig. \ref{color}], point sources removed, 
background from a 5 arcmin$^2$ region $\sim$8\arcmin\ NNE of the nucleus
outside the X-ray halo detected by ROSAT)  
     }
    \label{spec_epic}
\end{figure*}

The EPIC data clearly reveal more point sources than the deep ROSAT
observations and allow the mapping of diffuse emission in the disk (Fig.
\ref{color}). The underlying surface brightness at an energy of $\sim$1 keV
increases by factors of $\sim$10 between disk, plume and extended nucleus.
In the following we concentrate on the ten brightest \ngc253 
sources and present a brief analysis of the diffuse emission in the disk.

\subsection{Point sources}
   \begin{table}
      \caption{Bright point-like sources within \ngc253: EPIC (MOS1+2 \& PN) count rates, hardness ratios 
(with errors on the last digit[s]) 
 and luminosities in 10$^{37}$~erg~s$^{-1}$ both for EPIC (B) broad (0.2$-$10\,keV) 
band and for comparison (S) ROSAT (0.2$-$2.4\,keV) band (assumed \ngc253 distance: 2.58\,Mpc). }
         \label{sources}
         \begin{flushleft}
         \begin{tabular}{lrrrrrrr}
            \hline
            \noalign{\smallskip}
$\!\!\!$Src.	&\multicolumn{2}{c}{Count rate}&\multicolumn{3}{c}{EPIC hardness ratios}
   &\multicolumn{2}{l}{$L_{\rm x}^{\rm B}$$\ \ L_{\rm x}^{\rm S}$} \\
        &\multicolumn{2}{c}{MOS1+2 $\;\;$ PN} 
    &\multicolumn{3}{c}{HR1 $\;\;\;$ HR2 $\;\;\;$ HR3}& \multicolumn{2}{l}{(10$^{37}$}\\
        &\multicolumn{2}{c}{(ct ks$^{-1}$)}& & & &\multicolumn{2}{r}{erg~s$^{-1}$)} \\
%        \multicolumn{2}{c}{($^{\dagger}$)$\;\;$($^{\dagger}$)} \\
            \noalign{\smallskip}
            \hline
            \noalign{\smallskip}
$\!\!\!$X34    &\multicolumn{2}{r}{445(23) 470(24)}  &\multicolumn{3}{r}{1.00(0) -0.49(1) -0.40(1)} &\multicolumn{2}{r}{120\ \ \ 49}\\
$\!\!\!$X33    &\multicolumn{2}{r}{303(15) 293(15)}  &\multicolumn{3}{r}{1.00(0) -0.12(1) -0.57(1)} &\multicolumn{2}{r}{95\ \ \ 32}\\
%$\!\!\!$X21    &\multicolumn{2}{l}{169($\;\,$4)}         &\multicolumn{3}{r}{0.85(1) -0.40(1) -0.66(2)} &\multicolumn{2}{r}{50\ \ \ 20}\\
$\!\!\!$X21    &\multicolumn{2}{l}{169($\;\,$4)}          & & & &\multicolumn{2}{r}{50\ \ \ 20}\\
$\!\!\!$X36    &\multicolumn{2}{r}{ 86($\;\,$2) $\!\:$111($\;\,$3)} &\multicolumn{3}{r}{0.94(1) -0.39(1) -0.75(2)} &\multicolumn{2}{r}{23\ \ 9.5}\\
$\!\!\!$X12    &\multicolumn{2}{r}{ 81($\;\,$2) $\;\;$99($\;\,$3)} &\multicolumn{3}{r}{0.97(1) -0.16(2) -0.67(2)} &\multicolumn{2}{r}{25\ \ 7.5} \\
$\!\!\!$T      &\multicolumn{2}{r}{ 28($\;\,$2) $\;\;$26($\;\,$2)} &\multicolumn{3}{r}{0.88(2) -0.10(3) -0.33(3)} &\multicolumn{2}{r}{14\ \ 2.1}\\
$\!\!\!$X40    &\multicolumn{2}{l}{$\;\;$27($\;\,$1)}     & & & &\multicolumn{2}{r}{9.2\ \ 2.8}\\
$\!\!\!$X19    &\multicolumn{2}{r}{ 19($\;\,$1) $\;\;$26($\;\,$1)} &\multicolumn{3}{r}{0.80(3) -0.41(3) -0.46(5)} &\multicolumn{2}{r}{5.5\ \ 1.8}\\
$\!\!\!$X28    &\multicolumn{2}{r}{ 17($\;\,$1) $\;\;$24($\;\,$1)} &\multicolumn{3}{r}{0.86(2) -0.54(2) -0.61(5)} &\multicolumn{2}{r}{4.7\ \ 1.5}\\
$\!\!\!$X23    &\multicolumn{2}{r}{ 17($\;\,$1) $\;\;$22($\;\,$1)} &\multicolumn{3}{r}{0.53(3) -0.57(3) -0.47(7)} &\multicolumn{2}{r}{4.6\ \ 2.3}\\
\noalign{\smallskip}
\hline
         \end{tabular}
         \end{flushleft}
\vskip-.5cm
   \end{table}
\begin{figure}
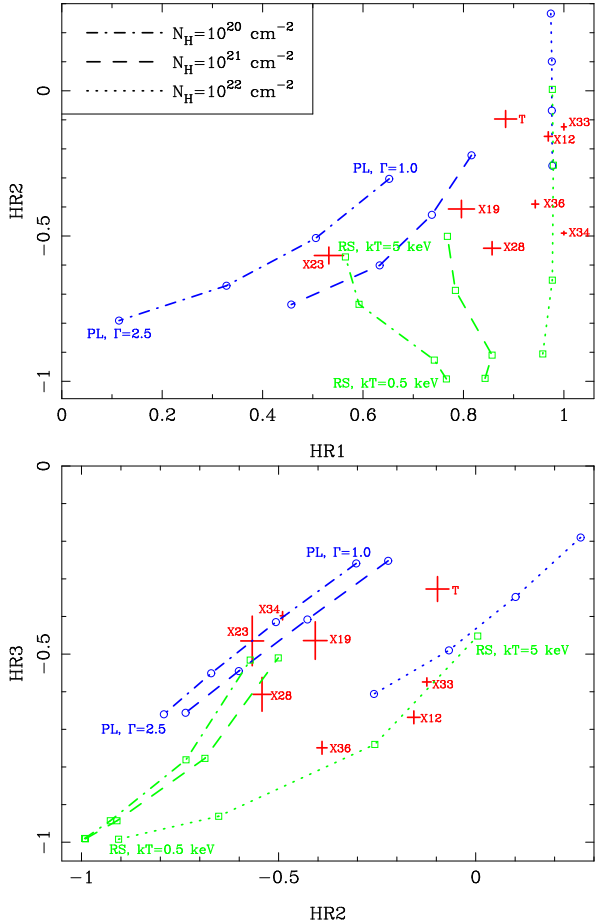

  \resizebox{7.8cm}{!}{\includegraphics[angle=-90,clip]{XMM18f3a.eps}}
%  \vspace*{1mm}
  \resizebox{7.8cm}{!}{\includegraphics[angle=-90,clip]{XMM18f3b.eps}}
    \caption[]{
    Colour-colour plots of HR2 versus HR1 (top) and
    HR3 versus HR2 (bottom) of sources from Table \ref{sources} (labeled
    red crosses indicating error range). 
    Positions of power law spectra (blue, photon index 1.0, 1.5, 2.0, 2.5
    marked by open circles) and thin, solar abundance thermal plasmas (green,
    temperatures of (0.5, 1.0, 2.0, 5.0) keV marked by open squares) are shown,
    connected up for absorbing columns of (0.1, 1.0, 10)$\times 10^{21}$ cm$^{-2}$
    }
    \label{hrs}
\end{figure}

We searched for the ten brightest 
on-galaxy sources detected with the highest maximum likelihood in the summed
low background MOS1 and MOS2 images to find new
\ngc253 X-ray transients (Table \ref{sources}). Where available
we use the source numbers introduced for the ROSAT \ngc253 point-like sources
(Vogler \& Pietsch \cite{vog99}). This list was confirmed by a visual inspection,
and source X28 was added (detected by PN only, though obviously real). 
For two sources (X21 and X40) the PN count rates and hardness ratios 
are not given as they lie close to the chip gaps.
The sources X17/X18 (two adjacent boxed sources north of X19 in Fig. \ref{color}) could 
not be distinguished by the SAS detection algorithms, and hence will be 
discussed in a later full point source analysis of \ngc253. 

Count rates were calculated using the total source count output 
in the (0.2--10)~keV band from the SAS source detection software. Due to
the complex background,  the 4 sources closest to the nucleus 
(X34, X33, X28 and T) were treated individually. We assumed a 
systematic error for the count rates of  5\% for the 4 
near-nuclear sources and 2\% for the other sources which
dominates the overall error. Hardness ratios (HRs) were determined  
from  4 bands ( [0.2--0.5]~keV, [0.5--2.0]~keV, [2.0--4.5]~keV, 
[4.5--10.0]~keV) by aperture photometry to get raw counts per source per
band on combined low background images from all the detectors
and exposures. The HRs have been calculated as hard -- soft / hard + soft,
with HR1 being (0.2--0.5)~keV vs (0.5--2)~keV and so on ('soft' here means the
count rate in the [0.2--0.5]~keV band etc.) with errors as
per Appendix A of Ciliegi et al. (\cite{ciliegi}). 
In Fig. \ref{hrs}  we show the position of the sources in HR colour-colour plots,
where we compare them to absorbed power law and absorbed, solar abundance
thin thermal plasma models.  The model points have been calculated using
the observation-specific combinations of EPIC responses, filters and
integration times. 

All bright persistent sources known from the ROSAT observations are also
detected by XMM-Newton. While the EPIC luminosity of most of the sources is
similar to that measured by ROSAT, X21 is brighter by a factor of two and X40
fainter by more than a factor of 3. In addition we detect 
$\sim70\arcsec$\ SSW of the nucleus one new bright
transient source (T, $\alpha_{2000}= 0^{\rm h}~47^{\rm m}~30\fs9$,
$\delta_{2000}=-25\degr 18\arcmin 26''$) 
that already was visible during the \chandra\ observation half a year earlier
(Strickland et al. \cite{str00}). HR1 is mainly an indicator of the absorption depth
within \ngc253 under which a source is seen, while HR2 and HR3 further 
characterize the spectrum. 
The positions of most of the sources in the two colour-colour plots are consistent
with a spectrum dominated by a single component. However, the nuclear source 
(X34) seems to be highly absorbed in HR1--HR2 and the opposite in HR2--HR3 
indicating a complex spectrum, as confirmed by the modeling in Sect. \ref{s41}.
The transient (T) is clearly the hardest source in the sample and highly absorbed. 
X33 and X12 seem to be even more absorbed but have a softer spectrum. 
Their high X-ray luminosity may
indicate that these sources are BHXRBs and the transient may be an
X-ray nova such as have been seen in our Galaxy (Tanaka \& Shibazaki \cite{tan96}). 
X19, X23, and X28 are less
absorbed and also have softer spectra  indicative of low mass X-ray binaries (LMXB). 
Most of the sources show time variability during the ROSAT and/or XMM-Newton
observations, arguing for a LMXB nature. The low HR3 of X36 together with the 
lack of time variability may point to it being a SNR.

\begin{figure}
  \resizebox{8.0cm}{!}{\includegraphics[angle=-90,clip=]{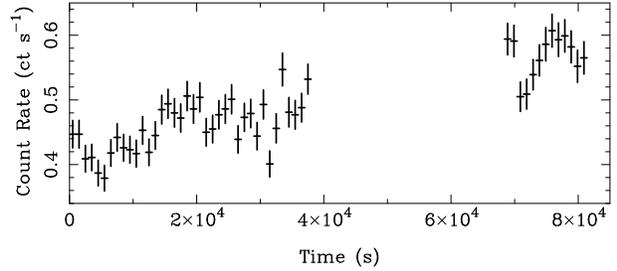}}
    \caption[]{ 
     X33 EPIC light curve. Data from all EPIC instruments in the (0.5--10)~keV
     band are integrated over 1000 s bins 
     }
    \label{x33_lc}
\end{figure}

While a detailed spectral and time variability study of all the \ngc253
sources is beyond the scope of the present paper, we demonstrate here 
the capabilities of the EPIC instrument on the brightest point
source, X33. The source (as well as X21) varies by a
factor $\sim 2$ during the XMM-Newton observation (see Fig.~\ref{x33_lc}),
however no significant spectral variability is detected. 
Its spectrum (Fig. \ref{spec_epic}) 
can be fit by a thermal bremsstrahlung model, assuming an
absorbing column of $N_{\rm H}$ = ($2.50\pm0.17$)$\times 10^{21}$\ cm$^{-2}$ and
temperature of ($5.3\pm0.5$)~keV ($\chi^2_{\rm red}$ = 0.98 for 180 d.o.f.).
The prefered model for BHXRBs (a disk-blackbody plus power law, see e.g.
Makishima et al. \cite{mak86}) gives an excellent fit ($N_{\rm H}$ = 
($5.1^{+2.6}_{-1.7}$)$\times 10^{21}$\ cm$^{-2}$, k$T$ = $1.55^{+0.14}_{-0.17}$~keV,
$r_{\rm in}$(cos $i$)$^{1/2}$ = $10.2^{+3.0}_{-2.1}$\ km assuming \ngc253 distance, 
and photon index of $4.3^{+1.7}_{-2.1}$, $\chi^2_{\rm red}$ =
0.88 for 178 d.o.f.). The high temperature and low $r_{\rm in}$(cos $i$)$^{1/2}$
when compared to Galactic BHXRBs may be explained by the black hole rotation,
which makes the disk get closer to the black hole and hence hotter, as suggested for
other ultraluminous compact X-ray sources in nearby spiral galaxies by Makishima
et al. (\cite{mak00}).

\subsection{Diffuse disk emission}
\label{s32}
The EPIC image reveals unresolved emission from the inner disk that is harder
along the inner spiral arms. We extracted EPIC PN spectra NE of the
nucleus, selecting areas of harder emission close to the major axis
(N) and softer emission adjacent to  the S
(yellow greenish and red; see Fig.~\ref{color}).  
While the
spectra below 0.5~keV are very similar (Fig. \ref{spec_epic}), the NE(N) shows
additional emission extending to energies of $\sim$2~keV.
They both indicate emission lines from \ion{O}{vii} and \ion{Fe}{xvii} 
pointing at major hot plasma origin. We therefore modeled the spectra,
assuming ISM components of solar abundance  with two temperatures,
shining through
the ISM, i.e. correcting for the Galactic foreground of
$N_{\rm H}$ = 0.13$\times 10^{21}$\ cm$^{-2}$ (Dickey \& Lockman 1990) and adding 
additional absorption within \ngc253.  For the NE(N) spectrum, we had to 
introduce a power law component which may describe contributions from unresolved 
point sources. Without including systematic errors that one might expect
from calibration uncertainties for the large extraction areas, we get
reasonable spectral fits to the (0.2--2.5)~keV band ($\chi^2_{\rm red}$ = 
1.5 and 1.6 for 30 and 17 d.o.f. for NE(N) and NE(S), respectively). 
The derived distinct lower and higher temperatures (0.13~keV and 0.5~keV) 
agree in both areas within the errors, and the
cooler component does not need additional absorption within \ngc253, 
indicating that the emission originates from the halo above the disk.
These temperatures, derived over small areas, agree with the ROSAT
results derived for the entire disk. More detailed analysis, including all EPIC 
detectors and exposures, will lead to a temperature map of the \ngc253 disk.

\section{Emission from the nuclear region}

\begin{figure}
  \resizebox{8.0cm}{!}{\includegraphics[angle=-90,clip=]{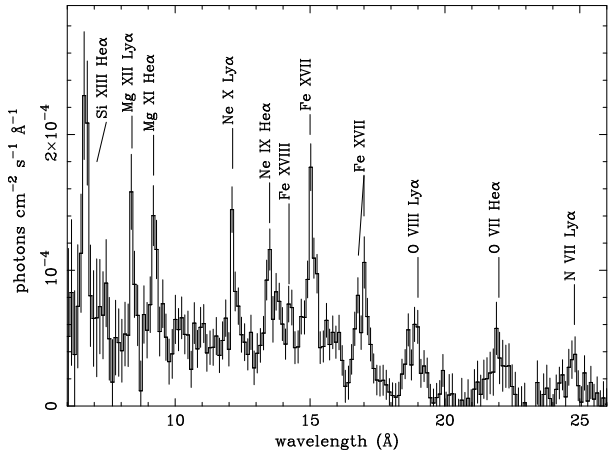}}
    \caption[]{ 
     ``Fluxed" RGS spectrum of the bright nuclear area of \ngc253 (extraction 
     region 1' along the minor disk axis, covering nucleus and plume). Bright
     emission lines are identified
     }
    \label{spec_rgs}
\end{figure}

The nuclear region of \ngc253 (see inlay Fig. \ref{color}) is bright enough
for a detailed study with both RGS and EPIC.   
The RGS spectrum (Fig. \ref{spec_rgs}) is dominated by emission
lines of hydrogenic and heliumlike charge states of the abundant low
Z elements (N, O, Ne, Mg, Si) and the neonlike and fluorinelike charge
states of Fe. With the help of EPIC we can further localize the RGS emission 
components in the nuclear area (e.g. Fig. \ref{lines}).  
We have not performed quantitative spectral fits to the RGS
data, but several conclusions can be drawn directly from 
Fig. \ref{spec_rgs}.  
The strength of the Fe L lines, specifically the \ion{Fe}{xvii} line
at 15 \AA, relative to the K-shell lines suggests that collisional
ionization is the dominant soft X-ray emission mechanism.  We
find no evidence of recombination emission in the RGS band.  The
inferred temperature ranges 
from $\sim$300 eV (\ion{O}{vii}, \ion{Fe}{xvii}) to $\sim$1.5~keV 
(\ion{Mg}{xii}, \ion{Si}{xiii}).  
The weakness of the longer wavelength lines suggests 
significant photoelectric absorption with implied column densities in 
the range 10$^{21}$--10$^{22}$\ cm$^{-2}$.  The abundances do not appear to be 
unusual, although accurate abundance estimates will depend on the precise 
temperature distribution, which has yet to be determined.  The general 
appearance of the spectrum is reminiscent of the spectrum of intermediate 
age supernova remnant gas, as might be expected for a starburst nucleus
and the interaction of the outflowing wind with the cooler gas of the ISM
in the plume.  
The characteristic emission measure in this region is 
$\sim7.5\times 10^{61}$\ cm$^{-3}$ 
(assuming solar abundances), which, given the 30\arcsec\ extent 
of the nuclear source, implies a characteristic electron density 
of $\sim0.1$~cm$^{-3}$.

\subsection{Unresolved X-ray nucleus}
\label{s41}
\begin{figure}
  \resizebox{8.5cm}{!}{\includegraphics[clip]{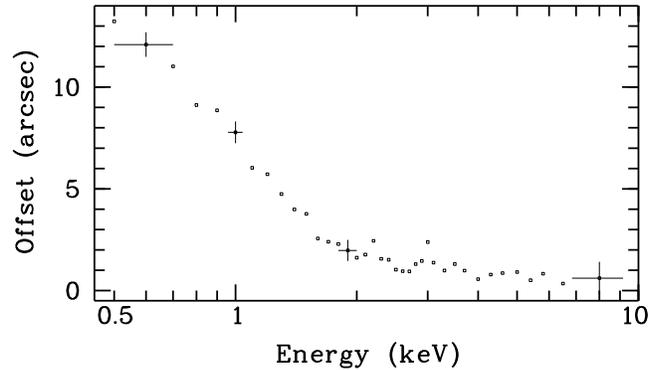}}
    \caption[]{
     Distance of the center of the X-ray emission from the position of the 
     galaxy nucleus 
     as a function of energy. The position in RA and Dec has been determined
     using at least 300 photons per energy bin in a radius of 12\farcs5.
     There is a strong shift along the minor axis of 10\arcsec\
     from 0.5~keV to 1.5~keV and a smaller shift up to 4~keV. Typical errors are 
     indicated 
     }
    \label{nuc_pos}
\end{figure}

Below 0.5~keV, no significant emission from the unresolved nuclear source 
(X34) can be seen. As
we go to higher energies, the centroid of the X34 emission shifts to the NW
along the minor axis (Fig. \ref{nuc_pos}) towards the position of the radio
nucleus indicating that only at energies above 4~keV will emission from the nuclear
starburst dominate the X-ray spectrum. 
It is apparent also in the inlay of Fig.\,\ref{color}, that harder (bluer) 
emission is shifted more to the NW where the black contours of the (2--10)~keV emission
are centered on the radio nucleus. The natural explanation is that 
X34 is an unresolved source of similar size to the shift reaching out from the
starburst nucleus (i.e. the collection of sources and diffuse emission as seen
with \chandra), that is 
increasingly absorbed towards the galaxy centre in the intervening
inner disk of \ngc253. 

Strickland et al. (\cite{str00}) report that the \chandra\ spectrum of the diffuse emission 
of the nuclear area (their NC) consists of a thermal plasma of temperature 0.66~keV,
4.7$\times 10^{21}$\ cm$^{-2}$ absorption plus emission in the (2--8)~keV band dominated
by point sources. The EPIC spectrum further characterizes the nuclear emission (Fig.
\ref{spec_epic}) without spatially resolving the \chandra\ point sources. 
While the spectral resolution is not sufficient to resolve the lines
seen by the RGS below 1~keV, lines from \ion{Mg}{xi}, \ion{Mg}{xii},
\ion{Si}{xiii}, \ion{Si}{xiv} \ion{Ar}{xvi}, and \ion{Fe}{xxv} are clearly
detected and argue for gas components with temperatures up to 5~keV and above. 
We modeled
the spectrum using thin thermal plasma components of solar abundance with corresponding
absorption increasing with the temperature of the plasma component and added a
power law component with the lowest absorption value. A model with three
temperature components gave an acceptable fit ($\chi^2_{\rm red}$ = 1.05 for
222 d.o.f.) with $N_{\rm H}$ = (0.34,1.78,13.2)$\times 10^{22}$ cm$^{-2}$, photon
index of 1.0, and temperatures of (0.56, 0.92, 6.3)~keV, respectively. 
A model with the same abundance in all three temperature components does not
require the power law component and gives for an abundance of 0.7 solar
an equally acceptable fit with very similar absorption values and temperatures.
The low temperature component is in good agreement with the diffuse \chandra\
emission. 
Note that within the nuclear spectrum, no evidence for a significant AGN 
contribution is detected (which would require a highly absorbed non-thermal component)
and the very hard component can therefore be attributed purely to
the starburst nucleus. The column density of $>10^{23}$\ cm$^{-2}$ is in good agreement
with the predictions from other wavelengths (see discussion in Pietsch et al. \cite{pie00}).

The energy, intensity and equivalent width of the Fe K line are 
(6.67$\pm$0.05)~keV, (7$\pm2)\times 10^{-7}$\ photons cm$^{-2}$s$^{-1}$ and
(930$\pm$300) eV.
Similar high-temperature plasma and Fe K lines have been 
found by XMM-Newton in young type\,\rm{I}b and type\,\rm{II}a SNRs, such as 
\casa (age 300\,yr) and \n132d (3000\,yr) (e.g. Behar et al. \cite{beh01}). 
Assuming that type\,\rm{II}a SNRs are prevalent within the starburst nucleus 
of \ngc253, and assuming the Behar et al. Fe line flux observed in 
\n132d, we estimate that $\sim$1000 such SNRs are necessary to 
explain the Fe K line intensity from \ngc253.
Furthermore, assuming that these SNRs
are strong iron line emitters for $\sim$5000\,yr, a supernova rate of
0.2\,yr$^{-1}$ is obtained, consistent with values estimated in other
wavebands (e.g. Ulvestad \& Antonucci \cite{ulv97}; Rieke et al. \cite{rie88}).

Some contribution to the hard X-rays from unresolved absorbed 
X-ray binaries, a low luminosity AGN and/or extended diffuse emission 
is not required by the data, but cannot be excluded. If present, this 
would lower the above estimates.
\subsection{X-ray plume emission}
\label{s42}
\begin{figure}
  \resizebox{8.0cm}{!}{\includegraphics[clip]{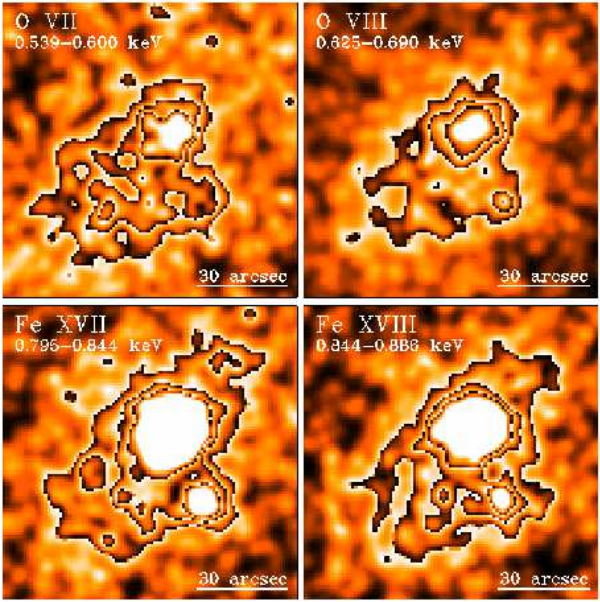}}
    \caption[]{
    XMM-Newton EPIC PN images of the \ngc253 
    nuclear area in the energy bands of four O and Fe
    emission lines formed 
    from low background times of both exposures using singles
    and doubles:({\bf upper left}) \ion{O}{vii} (20.7--23.0 \AA), 
                ({\bf upper right}) \ion{O}{viii} (18.0--19.6 \AA),
                ({\bf lower left}) \ion{Fe}{xvii} (14.7--15.6 \AA),
                ({\bf lower right}) \ion{Fe}{xviii} (14.0--14.7 \AA) 
    }
    \label{lines}
\end{figure}

The EPIC images (Fig. \ref{color}) trace the bright X-ray plume emission of
\ngc253 out to 1\farcm75 (1300 pc projected distance) from the nucleus along
the minor axis of the galaxy disk to the SE into the galaxy halo, much further
than possible with earlier observations. To test the energy dependence of the
emission we first selected images in broad energy bands. While the
limb-brightened structure was not visible below 0.5~keV, it clearly showed up
in the (0.5--0.9)~keV band. We also tried, using EPIC PN images, to spatially localize 
the dominant emission regions for the different ionization state ions seen with the
RGS. In both the
oxygen and the iron case (see Fig. \ref{lines}), while the low-ionization line
image shows the south-eastern plume to be quite uniformly filled, the
high-ionization line image shows definite limb-brightening, due to emission
predominantly from a shell embedded in the more uniformly filled plume region
in other energies. This clearly demonstrates the complexity of the emission
processes involved and good spectral fits on spatially integrated spectra can
not be expected. Nevertheless, we integrated a spectrum (Fig. 
\ref{spec_epic}) covering the region 
where Strickland et al. (\cite{str00}) derived an acceptable one temperature fit 
for \chandra\ (their CLB, k$T$ = 0.55~keV with absorption compatible to the Galactic foreground,
using subsolar abundances).
For EPIC PN, due to the broader energy response, the adding of two additional 
temperatures (k$T$ of [0.15, 0.53, 0.94]~keV and absorption within \ngc253 of
[0, 0, 1.3]$\times 10^{21}$\ cm$^{-2}$, using solar abundance) 
gave an acceptable fit ($\chi^2_{\rm red}$ = 
1.06 for 143 d.o.f.).  The soft component resembles the lower temperature of the
disk fits (see Sect. \ref{s32}) and the harder components the two softer temperatures
of the nuclear fit (see Sect. \ref{s41}). The fact that the harder components
seem to be less absorbed than in the nuclear spectrum, matches with the idea
that in the plume we are seeing out-flowing plasma mass-loaded close to the 
starburst nucleus. 

From the limb-brightened plume morphology seen by ROSAT and \chandra\ in overlays
on H$\alpha$ and in radial plots, Strickland et al. (\cite{str00}) come to the
conclusion that ``both the X-ray and H$\alpha$ emission come from low volume
filling factor gas, regions of interactions between a tenuous starburst-driven
wind of SN-ejecta and the dense ISM, and not from the wind itself". 
This explanation is in agreement with our results for the lines
of higher ionization states.  However, we observe more uniform
morphology for lower ionization states and below 0.5~keV, which
suggests that this emission is from a more uniformly
distributed mass, shocked and ``loaded" by the wind (e.g. Suchkov et al.
\cite{suc96}).
More detailed
modeling will allow us to calculate mass loading factors by comparing the
nuclear SNR rate with the mass seen in the plume.

\begin{acknowledgements}
    We thank the referee Guiseppina Fabbiano for her comments that helped to
    improve the manuscript considerably.
    The XMM-Newton project is supported by the Bundesministerium f\"{u}r
    Bildung und Forschung / Deutsches Zentrum f\"{u}r Luft- und Raumfahrt 
    (BMBF/DLR), the Max-Planck Society and the Heidenhain-Stiftung.
\end{acknowledgements}

\end{document}